\documentclass[sigconf]{acmart}

\AtBeginDocument{%
  }

\usepackage[linesnumbered,ruled,vlined]{algorithm2e}
\begin{document}

\title{FinRLlama: A Solution to LLM-Engineered Signals Challenge at FinRL Contest 2024}

\author{Arnav Grover}
\email{grover41@purdue.edu}
\orcid{0009-0004-9422-5426}
\affiliation{%
  \institution{Purdue University}
  \city{West Lafayette}
  \state{Indiana}
  \country{USA}
}

\begin{abstract}
     In response to Task II of the FinRL Challenge at ACM ICAIF 2024, this study proposes a novel prompt framework for fine-tuning large language models (LLM) with Reinforcement Learning from Market Feedback (RLMF). Our framework incorporates market-specific features and short-term price dynamics to generate more precise trading signals.  Traditional LLMs, while competent in sentiment analysis, lack contextual alignment for financial market applications. To bridge this gap, we fine-tune the LLaMA-3.2-3B-Instruct model using a custom RLMF prompt design that integrates historical market data and reward-based feedback. Our evaluation shows that this RLMF-tuned framework outperforms baseline methods in signal consistency and achieving tighter trading outcomes; awarded as winner of Task II. You can find the code for this project on \href{https://github.com/Arnav-Gr0ver/ICAIF_FinRL-2024}{\textcolor{blue}{GitHub}}.
\end{abstract}


\begin{teaserfigure}
  \centering
  \includegraphics[width=250pt]{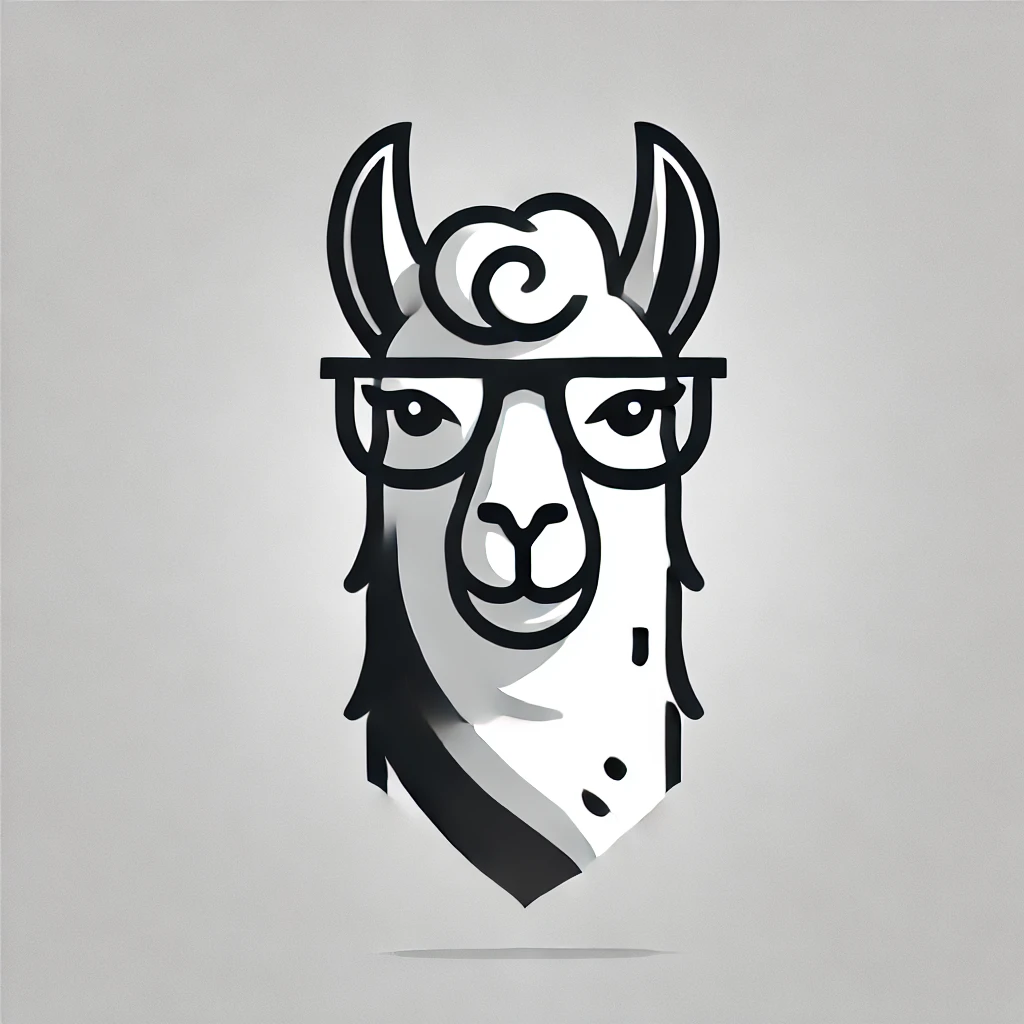}
  \captionsetup{font=large}
  \caption{FinRLlama}
  \label{fig:teaser}
\end{teaserfigure}

\maketitle

\section{Introduction}

The application of large language models (LLMs) to financial sentiment analysis represents a significant opportunity for algorithmic trading strategies \cite{Liu22}. Although LLMs demonstrate sophisticated language understanding capabilities, their application in financial contexts has been limited by the challenge of incorporating market-specific knowledge and temporal dynamics \cite{Nie24}.

\subsection{Background and Related Work}

The evolution of financial sentiment analysis has followed several key trajectories in the literature. Early work by Loughran and McDonald \cite{Loughran11} established the importance of domain-specific dictionaries for financial text analysis, highlighting how general-purpose sentiment tools often fail in financial contexts. In parallel, the development of comprehensive financial reinforcement learning frameworks like FinRL-Meta \cite{Liu20} has provided standardized environments for developing and evaluating trading strategies.

Recent developments in prompt engineering have shown promising results in various domains. Wei et al. \cite{Wei22} demonstrated how carefully constructed prompts can elicit domain-specific knowledge from LLMs without fine-tuning, while Vatsal and Dubey \cite{Vatsal23} provided various methods / frameworks for evaluating prompt effectiveness in various NLP tasks. However, applications in financial sentiment analysis have been limited, with most approaches focusing on model architecture modifications rather than prompt optimization.

\subsection{Research Objectives}

In response, we establish the following objectives for this task:

\begin{enumerate}
    \item Develop a novel prompt framework for sentiment analysis.
    \item Establish a systematic training methodology for model optimization that adapts dynamically to market conditions.
    \item Empirically validate the framework’s impact on sentiment-based signal precision and trading performance.
\end{enumerate}

These objectives aim to create a robust framework for generating actionable insights from financial news, advancing the utility of LLM in financial applications through a combination of prompt engineering and market-aligned learning.

\section{Methodology}

\subsection{Prompt Architecture}

The sentiment analysis prompt architecture generates stock performance predictions from news headlines using a scalable sentiment framework. Building on established sentiment analysis methods in finance \cite{Loughran11}, \cite{Bollen11}, the system integrates market feedback and historical data to improve predictive precision \cite{MDPI23}. Adjustable parameters enables market adaptation, while single-score output supports rapid trading decisions.

\begin{algorithm}
\caption{Sentiment Signal Scoring Prompt}
\KwIn{Signal Bound, Threshold, News Headline, Price Data}
\KwOut{Value in [-\texttt{signal\_strength}, \texttt{signal\_strength}]}

\BlankLine

\textbf{[CONTEXT]} \\
Task: Analyze the stock-related news headline and output a sentiment score reflecting the sentiment's potential impact on stock performance.

\BlankLine

\textbf{[SENTIMENT SCORING PARAMETERS]} \\
\texttt{-signal\_strength:} Highly negative market sentiment \\
\texttt{-threshold:} Moderately negative market sentiment \\
\texttt{0:} Neutral market sentiment \\
\texttt{threshold:} Moderately positive sentiment\\
\texttt{signal\_strength:} Highly positive sentiment

\BlankLine

\textbf{[MARKET FEEDBACK CONSIDERATIONS]} \\
\texttt{Past Market Responses:} Incorporate past market responses to similar news events. \\
\texttt{Market Sentiment Alignment:} Evaluate if the news aligns with or contradicts prevailing market sentiment. \\
\texttt{Historical Price Patterns:} Analyze the historical impact of similar news on stock prices.

\BlankLine

\textbf{[SENTIMENT SCORING EXAMPLES]} \\
"Company X announces layoffs amid economic downturn." Sentiment Score: -8 \\
"Company Y reports record revenue growth in Q1." Sentiment Score: 7 \\
"Market responds positively to Company Z’s new product launch." Sentiment Score: 5

\BlankLine

\textbf{[OUTPUT]} \\
Integer sentiment score in range [\texttt{-signal\_strength}, \texttt{signal\_strength}] based on analysis.

\end{algorithm}

\subsection{Training Process}

\begin{figure}[h]
  \centering
  \includegraphics[width=\linewidth]{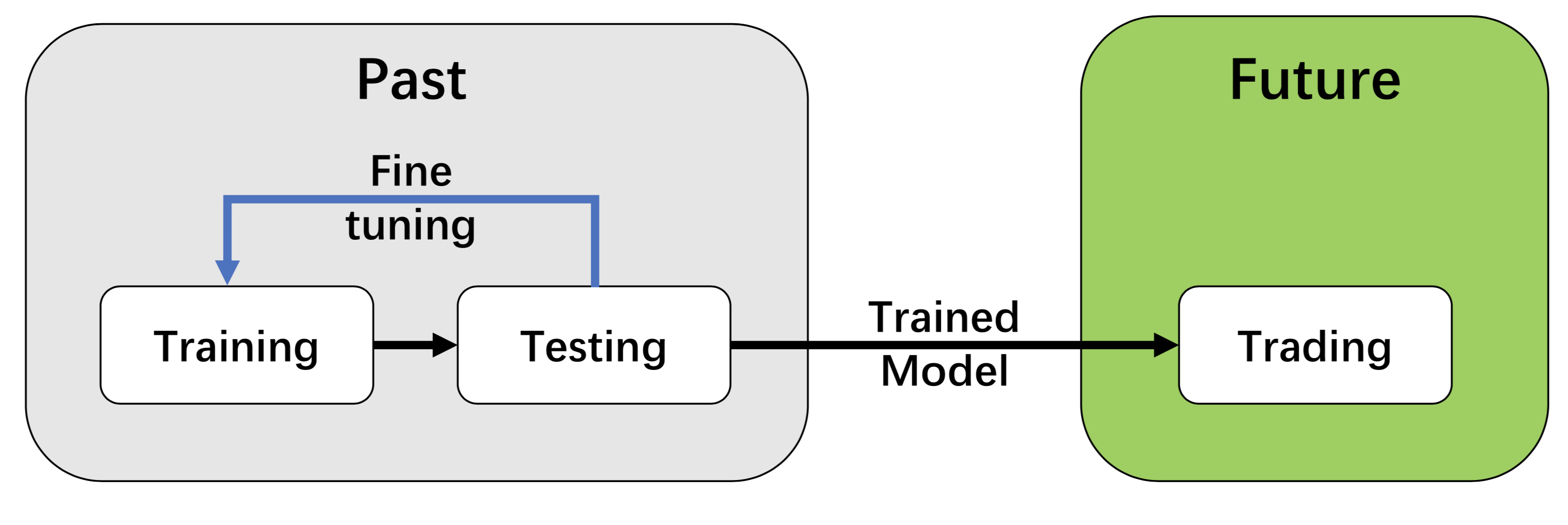}
  \captionsetup{font=large}
  \caption{FinRL Train-Test-Trade Pipeline}
\end{figure}

The model fine-tuning process begins with the base Llama-3.2-3B-Instruct model. The reinforcement learning (RL) component simulates market interactions, where the model outputs sentiment signals, and the model selects trading actions (long, short, or hold). The reward function then evaluates the model's predictions by comparing sentiment scores to actual market performance, assigning rewards or penalties based on the accuracy of the sentiment and resultant profits or losses. This process aligns with established RL frameworks in financial applications \cite{Wang23}, \cite{Mnih15}, and follows the task requirements outlined in the ACM ICAIF FinRL 2024 Competition \cite{Holzer24}.

The reward function is dynamically adjusted on the basis of the strength of the model's sentiment signal, reinforcing correct predictions, and penalizing errors. The function takes into account the confidence of the model, incorporating adjustable thresholds to assess market direction. For instance, when the sentiment score exceeds a threshold, the reward varies depending on the actual price movement: long positions are rewarded if a strong positive return is observed, while negative returns despite positive sentiment lead to penalties. Similarly, short positions are rewarded when negative returns align with the sentiment. This system helps the model refine its decision making over time through feedback loops, gradually improving its accuracy and trading strategies.

The model's fine-tuning process is guided by the Adam optimizer, minimizing the loss function based on the discrepancy between predicted sentiment signals and actual market outcomes. This approach follows deep-RL strategies aimed at optimal decision-making, balancing exploration, and exploitation to generate robust sentiment-based trading signals.

\section{Results}

\subsection{Experimental Setup}

The experimental setup for testing and validating the proposed model's performance spans 2020 to 2023, assessing the accuracy and profitability of sentiment-based trading signals against the baseline. This setup includes news headlines, stock price data, and technical indicators to effectively align sentiment scores with stock movements. Each headline is pre-processed to link with relevant stock price data, and a three-day forward close price is added to facilitate forward-looking impact analysis.

We designate 2020–2022 as the training period and use 2023 exclusively for evaluation. This split enables an assessment of the robustness of the model in diverse market conditions. For both the model and the baseline, each headline generates a buy, hold, or sell signal, with performance measured by cumulative returns, win/loss rate, etc.

\subsection{Performance Metrics}

\begin{figure}[h]
  \centering
  \includegraphics[width=\linewidth]{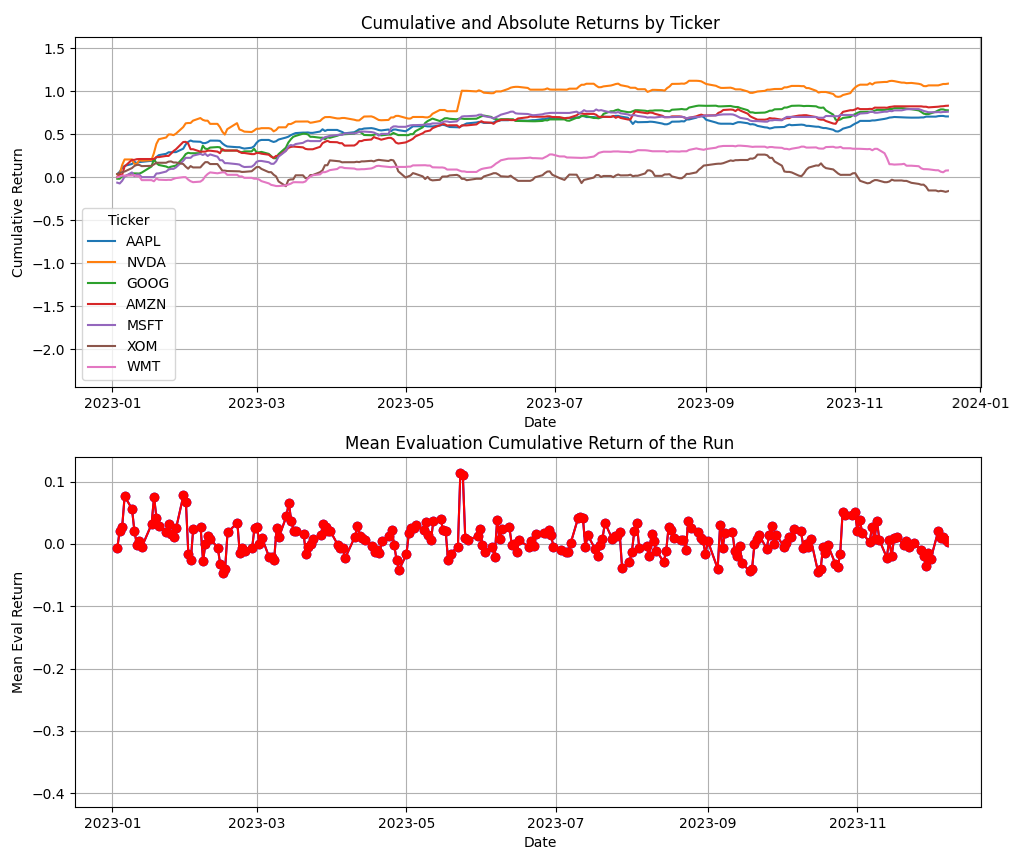}
  \captionsetup{font=large}
  \caption{FinRLlama Cumulative Returns}
\end{figure}

\begin{figure}[h]
  \centering
  \includegraphics[width=\linewidth]{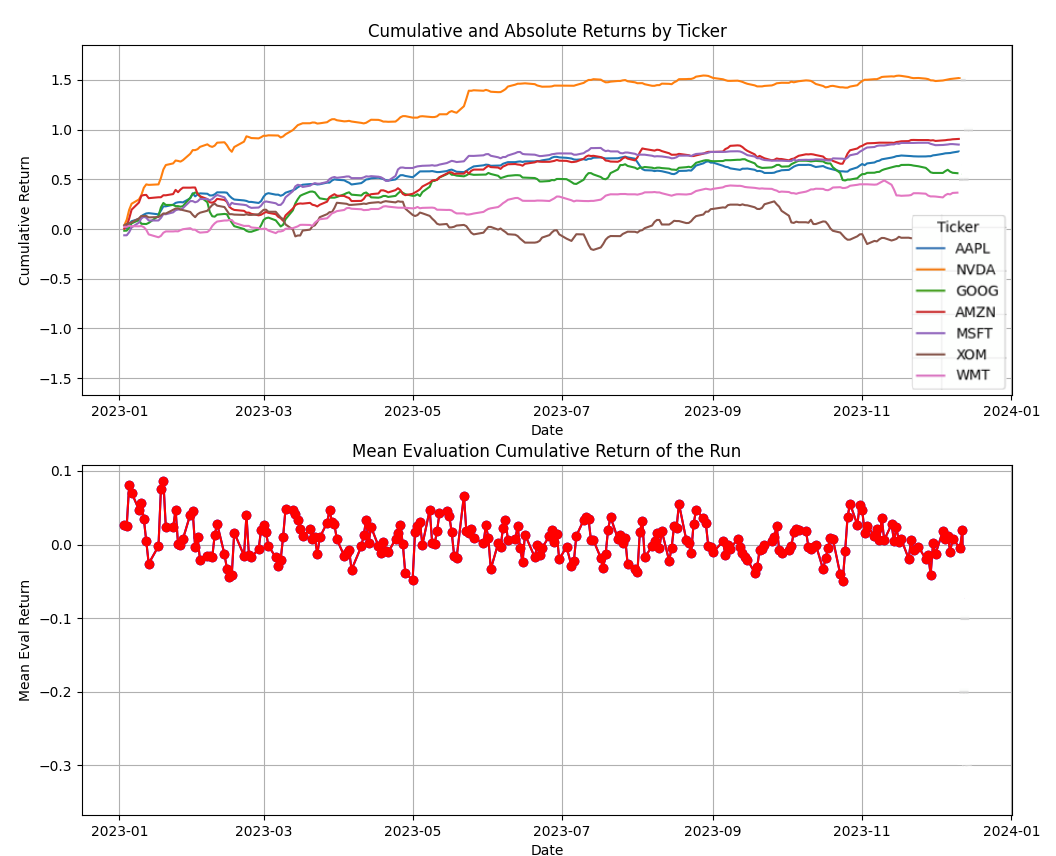}
  \captionsetup{font=large}
  \caption{Llama Cumulative Returns}
\end{figure}

\subsection{Comparative Analysis}

In Figure 3, the cumulative returns of the tickers appear to be less volatile. Although NVDA still leads with positive cumulative returns, its gains are not as pronounced and the spread between the highest and lowest performing stocks is narrower compared to the first graph. XOM continues to trend downward, but with a less steep decline. This indicates a model response that may be more conservative, possibly due to model fine-tuning. In Figure 3, the cumulative returns show significant variation between tickers. For example, NVDA displays consistently higher returns, reaching above 1.5, indicating strong performance. Other stocks like MSFT and GOOG exhibit moderate cumulative returns, staying close to 0.5, while XOM shows a downward trend - temporarily dipping into negative returns. This plot suggests a broader divergence in performance between stocks, with the model interpreting NVDA as markedly outperforming others and XOM underperforming.

The mean cumulative evaluation return, shown in both second subplots, oscillates around zero in both cases.

In summary, Llama-3.2-3B-Instruct displays a broader range of cumulative returns, indicating higher variability and greater individual gains and losses, while FinRLlama suggests a more conservative approach with reduced volatility in cumulative returns across tickers and smoother mean evaluation. 

\section{Future Work}

Future work to improve the model could focus on refining the reward function to better capture the nuances of financial market dynamics. By incorporating dynamic reward adjustments that account for market volatility and shifts in sentiment, the model could become more responsive to short-term fluctuations and long-term trends. Enhancing the model’s ability to process and integrate historical price data and sentiment trends could improve its prediction accuracy, allowing it to account for delayed market reactions more effectively. Furthermore, further fine-tuning with domain-specific financial data would help the model better adapt to the intricacies of market behavior, improving its decision-making accuracy. These improvements could significantly improve the robustness of the model and its ability to generate actionable trading insights.

\section{Acknowledgments}

Arnav Grover acknowledges the support of Keyi Wang, Nikolaus Holzer, and Xiao-Yang Liu Yanglet for their invaluable guidance and support throughout the development of this work. We also extend sincere thanks to Yanglet and the organizing team of the FinRL 2024 contest for their exceptional efforts in creating a collaborative and intellectually stimulating platform. Their dedication to fostering innovation and research in the field of open finance has provided an invaluable opportunity to contribute to this evolving domain.

\bibliographystyle{ACM-Reference-Format}

\end{document}